\documentclass[sigconf]{acmart}

\usepackage{fancyhdr}
\usepackage{url}
\usepackage{microtype}
\usepackage{algorithmic}
\usepackage{textcomp}
\usepackage{xcolor}
\usepackage{enumitem}
\usepackage{tikz}
\usepackage{siunitx}
\usepackage{soul}
\sethlcolor{yellow}
\usepackage{hyperref}

\usepackage{titlesec}
\titleformat*{\paragraph}{\bfseries}
\titlespacing*{\paragraph}{0pt}{1ex plus .1ex minus .2ex}{2pt}

\usepackage{setspace}
\usepackage{def}

\AtBeginDocument{%
  \providecommand\BibTeX{{%
    \normalfont B\kern-0.5em{\scshape i\kern-0.25em b}\kern-0.8em\TeX}}}

\begin{document}

\title[]{Building A Trusted Execution Environment \\ for In-Storage Computing}

\author{Yuqi Xue, Luyi Kang\footnotemark[2], Weiwei Jia, Xiaohao Wang, Jongryool Kim\footnotemark[3], Changhwan Youn\footnotemark[3], \\ Myeong Joon Kang\footnotemark[3], Hyung Jin Lim\footnotemark[3], Bruce Jacob\footnotemark[2], Jian Huang}

\affiliation{%
  \institution{UIUC, \footnotemark[2]University of Maryland, College Park, \footnotemark[3]SK Hynix}
  \streetaddress{Address}
  \city{}
  \state{}
  \country{}}

\renewcommand{\shortauthors}{} 





\maketitle

\section{Introduction}
\label{sec:intro}
In-storage computing has been a promising technique for accelerating data-intensive 
applications, especially for large-scale data processing and analytics~\cite{biscuit}. 
It moves computation closer to the data stored in the storage devices like  
flash-based solid-state drives (SSDs), such that it can overcome the  
I/O bottleneck by reducing the amount of data transferred between 
the host machine and storage devices. 
As modern SSDs are employing multiple general-purpose embedded processors 
and large DRAM in their controllers, it becomes feasible to enable in-storage 
computing in reality today. 

To facilitate the wide adoption of in-storage computing, a variety of frameworks have 
been proposed.
All these prior works show the great potential of in-storage 
computing for accelerating data processing in data centers. 
However, most of them focus on the performance and programmability, but few of them treat the security as the first 
citizen in their design and implementation, which imposes great threat to the user data and 
SSD devices, and further hinders its widespread adoption.

As in-storage processors operate independently from the host machine, and modern  
SSD controllers do not provide a trusted execution environment (TEE) for programs running inside the SSDs,  
they pose severe security threats to user data and flash chips. 
To be specific, a piece of offloaded (malicious) code
could (1) manipulate the mapping table in the flash translation layer (FTL) to mangle  
the data management of flash chips, (2) access and destroy data belonging 
to other applications, and (3) steal and modify the memory of co-located 
in-storage programs at runtime. Even worse, adversaries can steal and modify intermediate data and results generated by in-storage programs via physical attacks such as cold-boot attack, bus snooping attack, and replay attack~\cite{physicalattack:blackhat2014}.  

To overcome these security challenges, state-of-the-art in-storage computing frameworks maintain a copy of the privilege information in the SSD DRAM and enforcing permission checks for in-storage programs.
However, such a solution still suffers from many security vulnerabilities~\cite{sok}.
An alternative approach is to adopt Intel SGX.
Unfortunately, modern in-storage processors do not support SGX,
and it also incurs significant performance overhead~\cite{www:intelsgx}.

\begin{figure}[t]
\centering
\includegraphics[width=0.8\linewidth]{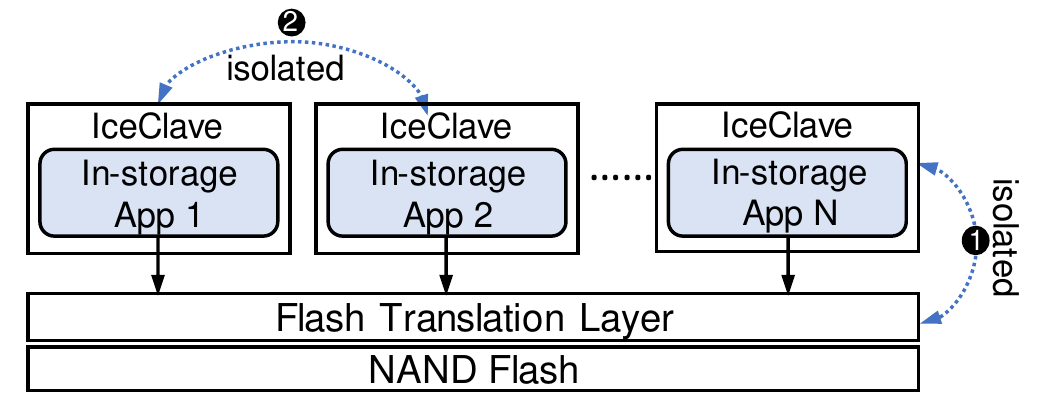}
\vspace{-1.5ex}
\caption{IceClave enables in-storage TEEs to achieve security isolation between in-storage programs, 
	FTL, and flash chips.}
\label{fig:goals}
\vspace{-3.5ex}
\end{figure}

Therefore, providing a secure, lightweight, and trusted execution environment for in-storage computing 
is an essential step towards its widespread adoption. 
Ideally, we wish to enjoy the performance benefits of in-storage computing, while 
enforcing the security isolation between in-storage programs, the core FTL functions, and physical 
flash chips, as demonstrated in Figure~\ref{fig:goals}.  

To this end, we present IceClave, a trusted execution environment for in-storage computing. 
IceClave is designed specifically for modern SSD controllers 
and in-storage programs, with considering the unique flash properties and in-storage workload characteristics. 
With ensuring the security isolation, IceClave includes (1) a new memory protection scheme to protect the FTL and reduce the context switch overhead 
incurred by flash address translations; (2) a technique for securing in-storage DRAM for in-storage programs by taking 
advantage of the fact that most in-storage applications are read intensive; (3) a stream 
cipher engine for securing data transfers between storage processors and flash chips, with low performance overhead and energy consumption; 
and (4) a runtime system for managing the life cycle of in-storage TEEs.

We implement IceClave with a full system simulator
and develop a real system prototype with a real-world OpenSSD Cosmos+ FPGA board.
Compared to state-of-the-art in-storage computing approaches,  
IceClave introduces only 7.6\% performance overhead to the in-storage runtime
and delivers 2.31$\times$ better performance than host-based computing,
while adding minimal area and energy overhead to the SSD controller. 
\section{\NoCaseChange{Threat Model}}
\label{sec:threatmodel}
We target the multitenancy where multiple application instances operate in the shared 
SSD. Following the threat models for cloud computing today,
we assume the cloud computing platform has provided a secure channel for end users to offload 
their programs to the shared SSD. The related code-offloading techniques, such as secure RPC and libraries, have already been deployed in cloud platforms~\cite{f1}. 
However, an offloaded program can include (hidden) malicious code.

We assume hardware vendors do not intentionally implant backdoor or malicious programs in their 
devices. However, as we deploy those computational SSDs in shared platforms (e.g., public cloud), 
we do not trust the platform operators who could initiate board-level physical attacks such as 
bus-snooping and man-in-the-middle attacks, or exploit the host machine to steal or destroy data 
stored in SSDs. Similar to the threat model for SGX, 
we assume that the processor chip is safe against physical attacks, and we exclude software side-channel attacks~\cite{www:intelsgx}.
\section{\mbox{\hspace{-0.89ex} Securing In-Storage Computing with IceClave}}
\label{sec:design}
IceClave is a TEE for in-storage computing with minimal performance and hardware cost. It aims to defend against three attacks: (1) the attack against co-located in-storage 
programs; (2) the attack against the FTL; (3) the potential physical attack 
against the data loaded from flash chips and generated by in-storage programs.

\paragraph{Protecting Flash Translation Layer.}
\label{subsec:ftl}
As the FTL manages flash blocks and controls how user data is mapped to each flash page, its protection is crucial. 
If any malicious in-storage programs gain control over it, they can read, erase, or overwrite data from other users, 
causing severe consequences such as data loss and leakage. 

To protect FTL from malicious in-storage programs, we have to guarantee offloaded applications 
cannot access memory regions used by FTL. We can use ARM TrustZone to create secure and normal worlds, and then place FTL functions in the secure world, and place all in-storage applications in the normal world. 
However, this will cause significant performance overhead for in-storage applications. 
This is because when an application accesses a flash page each time, it needs to context switch to the 
secure world which hosts the FTL and its address mapping table. 

To address this challenge,
we partition the entire physical main memory space into 
three memory regions: normal, protected, and secure by extending TrustZone.
We allow FTL to execute in the secure world, and place in-storage applications in the normal world; therefore, they cannot 
access any code or data regions that belong to the FTL. 
We use the protected memory region in the normal world to host the shared address mapping table, such that in-storage applications can only read the mapping
table entries for address translation, without paying the context-switch overhead.
\begin{figure}
	\centering
	\includegraphics[width=0.85\linewidth]{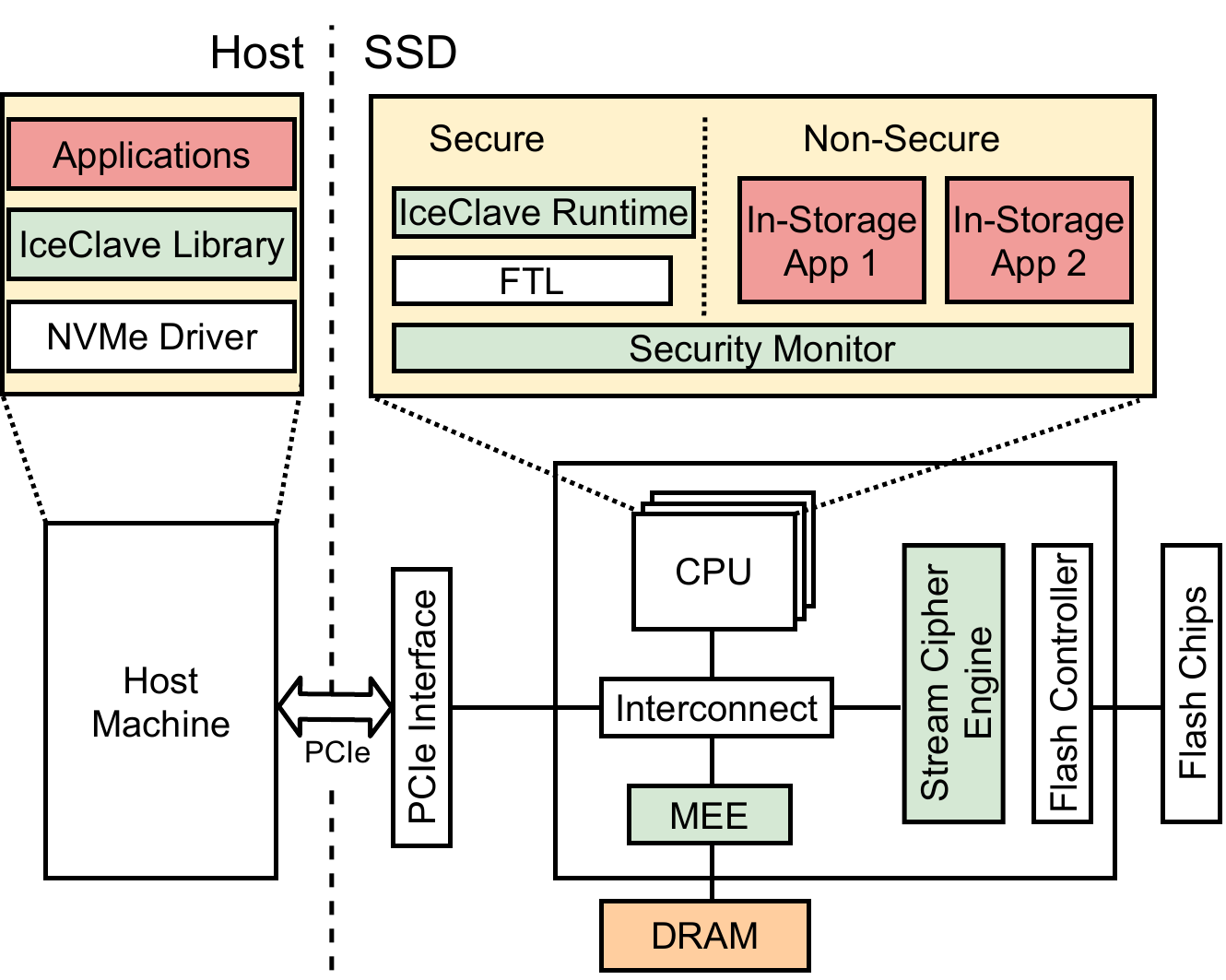}
	\vspace{-2ex}
	\caption{Overview of IceClave architecture.}
	\label{fig:overview}
\end{figure}
\paragraph{Enforcing Access Control for In-Storage Programs.}
\label{subsec:mapping}
Although each in-storage program only has the read access permission when accessing the mapping table of 
the FTL, a malicious in-storage program 
could probe the mapping table entries (e.g., by brute-force)
and easily access the data belonging to other in-storage programs.

To address this challenge, we extend the address mapping table of FTL.  
We use the ID bits in each entry (8 bytes per entry) to track the identification of each in-storage TEE, 
and use them to verify whether an in-storage TEE has the permission to access the mapping table entry or not. 
Each in-storage program only has accesses to the address mapping table of the FTL and allocated memory space. 
Accesses to other memory locations will result in a fault in the memory management unit.

\paragraph{Securing In-Storage DRAM.}
\label{subsec:memory}
In-storage programs load data from flash chips to the SSD DRAM for data processing. 
The user data that includes raw data, intermediate data, and produced results in the DRAM could be leaked or tampered with at runtime due to physical attacks. To address this challenge, IceClave enables both memory encryption and integrity verification. 

For memory encryption, the state-of-the-art work usually uses split-counter encryption~\cite{bmt:micro07}, which has significant performance overhead. However, this is less of a concern for in-storage computing because in-storage workloads are mostly read intensive. Based on this observation, we design the hybrid-counter scheme.

The key idea of hybrid-counter is that we only use major counters for read-only pages. For writable pages, we apply the traditional split-counter scheme.
As minor counters will not change as long as the pages are read-only, we do not need minor counters for read-only pages.
In this case, we can improve the counter fetching performance by packing more counters per cache line.

To ensure the processor receives exactly the same content as it wrote in 
the memory most recently, we also enable memory integrity verification by employing Bonsai Merkle Tree (BMT)~\cite{bmt:micro07}. Due to the hybrid-counter scheme, IceClave maintains two Merkle trees, but the extra memory cost is negligible.

\paragraph{IceClave Implementation.}
We show the overview of IceClave architecture in Figure~\ref{fig:overview}.
We extend ARM TrustZone to create secure and normal world for security isolation
and protection of different entities in FTL, while enabling memory
encryption and verification with memory encryption engine (MEE).
We implement IceClave with a computational SSD simulator developed based on the
SimpleSSD, Gem5, and USIMM simulator.
To verify the core functions of IceClave,
we also implement IceClave with an OpenSSD Cosmos+ FPGA board.

\begin{figure}[t]
    \centering
    \includegraphics[width=0.98\linewidth]{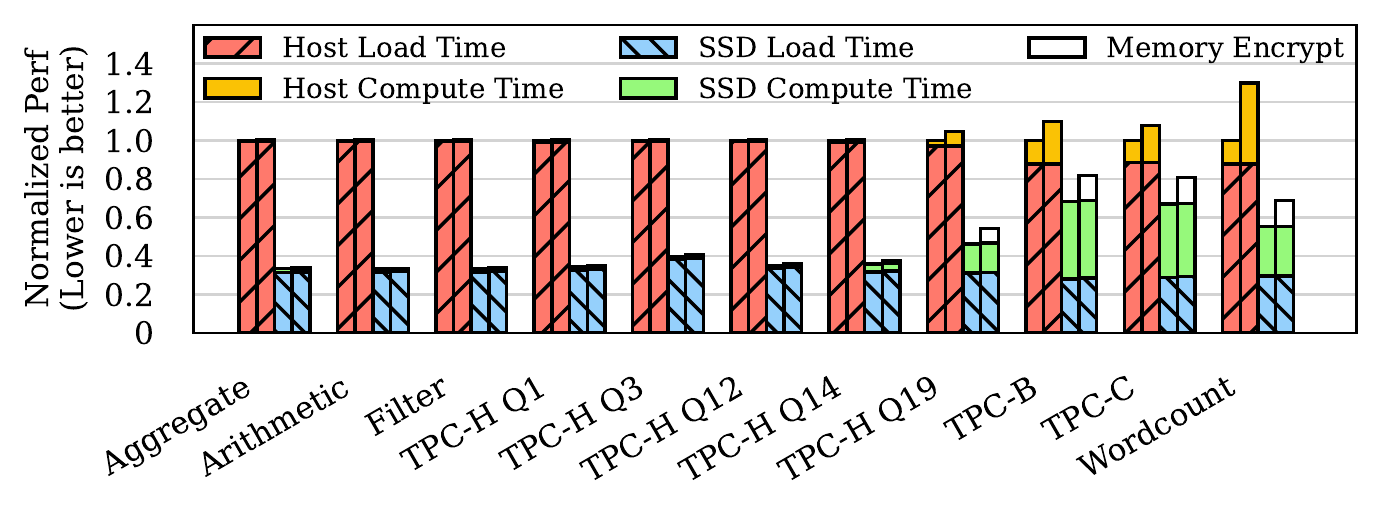}
    \vspace{-3ex}
    \caption{Performance comparison of Host, Host+SGX, ISC, and IceClave (from left to right).}
    \label{fig:iscvsvanila}
    \vspace{-3.5ex}
\end{figure}

\paragraph{Performance of IceClave.}
We evaluate IceClave with a set of synthetic and real-world workloads that are typical for in-storage computing.
We compare IceClave with the following state-of-the-art solutions: (1) host-based computing without security (\texttt{Host}), (2) host with Intel SGX (\texttt{Host+SGX}), and (3) in-storage computing without security (\texttt{ISC}).
As shown in Figure~\ref{fig:iscvsvanila}, IceClave outperforms 
Host and Host+SGX by more than 2.3$\times$, respectively.
Compared to the ISC baseline, IceClave introduces 
7.6\% performance overhead, due to the security techniques used in the in-storage TEE.
\section{Significance and Long-term Impact}
This work would have a long-term impact on
the future development and deployment of in-storage computing, and the future research on securing new computing paradigms.

\paragraph{(1) It provides a comprehensive threat analysis for in-storage computing.}
This paper is the first to focus on security threats faced by in-storage computing. 
We carefully investigated existing in-storage computing systems and discovered several vulnerabilities that may lead to severe consequences: (a) A malicious user can manipulate the intermediate data and output generated by in-storage programs via both software and physical attacks, causing incorrect computing results; (b) A malicious program can intercept FTL functions like GC and wear leveling in the SSD and mangle the flash management, causing data loss or device destroyed; (c) A malicious user can steal user data stored in flash chips via physical attacks.
%
We have to overcome these security challenges for its widespread deployment, considering they pose severe threats to user data and flash devices.

\paragraph{(2) It offers a security solution for in-storage computing.}
We investigated the feasibility of applying existing solutions to secure in-storage computing, unfortunately, 
most of them do not work properly and efficiently with in-storage computing.
For example, we can develop an OS, a hypervisor, or a SGX-like solution for in-storage computing. However, due to the limited resources  
in the SSD controller, they introduce significant overheads to the SSD and increase the attack surface, due to their large codebases. 

We proposed the first trusted in-storage computing framework that takes security as the priority. We demonstrated the necessity and feasibility of incorporating the unique flash properties and in-storage workload characteristics into the design for achieving strong security and decent performance at the same time. Our evaluation shows that the performance and hardware cost of IceClave are acceptable for real production systems. 
We believe that security will become a standard feature for computational storage devices, and storage vendors can use IceClave as a reference for building secure computational SSDs.

\paragraph{(3) It develops a new memory encryption and verification scheme.}
IceClave proposes to use hybrid encryption counters for different memory access patterns. While we apply this technique to exploit the fact that in-storage computing workloads are read-intensive, the same idea can be generalized to other systems if we know the workload characteristics. IceClave will inspire future work on memory encryption and verification schemes for applications with different memory access patterns.

\paragraph{(4) It has implications on new computing paradigms.}
As different computing paradigms (e.g., in-storage/in-memory computing, and accelerator-centric computing) offer different execution environments and have their unique architecture and security challenges, it is not easy to apply a TEE solution from one domain to another. This is also true as we shift from host-based computing to in-storage computing, especially with the goal of achieving both efficiency and enhanced security. We believe that IceClave will inspire the development of TEEs for these new computing paradigms. For example, we expect that a TEE framework would be built for securing accelerators. 

\paragraph{(5) It facilitates the deployment of computational storage in multi-tenant clouds.}
Computational storage devices are becoming more pervasive in clouds. 
Although TEEs have already been deployed for the host machines in the cloud, there is no support for in-storage TEEs, 
which poses significant security threats to sensitive user data.
IceClave is developed based on a realistic threat model for multi-tenant clouds and is well-suited for this scenario.

\paragraph{(6) It enables the secure deployment of computational storage in edge IoT systems.}
As computational storage is being widely deployed in edge IoT systems~\cite{aws_greengrass_ngd},
they suffer from physical attacks~\cite{iot_physical_attacks}. Since these devices are often deployed in the public or wild fields, it is extremely easy for an attacker to get physical access to IoT devices. The attacker can unsolder the device and steal data from the flash memory, and tamper with the storage controller to cause unintended behaviors.  
The limited hardware resource and power budget of edge devices introduce even more challenges for securing the computational storage.
IceClave is a lightweight solution that can protect computational storage against physical attacks, without introducing much overhead to the resource-limited edge devices.


\paragraph{(7) It provides the implementation flexibility for different processor architectures.}
IceClave leverages ARM TrustZone to enable the memory protection between 
in-storage programs and FTL functions. This is driven by the fact that ARM processors are available in a majority of 
modern SSD controllers. 
The key idea of IceClave can also be implemented with new type of processors. 
For instance, RISC-V defines three levels of privileges. We can map the normal, protected, and secure memory regions to different memory regions in RISC-V.
Beyond ARM and RISC-V processors, 
recent works also deployed hardware accelerators in SSD controllers~\cite{biscuit}. 
They are also lacking the in-storage TEEs. We believe IceClave will also facilitate the TEE development for in-storage accelerators.



\bibliographystyle{ACM-Reference-Format}
\bibliography{references, ref, ref_toppicks2021}

\end{document}